\documentclass[11pt,a4paper]{article}
\pdfoutput=1
\usepackage{jheppub}
\usepackage[T1]{fontenc} 

\def\beq{\begin{equation}}
\def\bea{\begin{eqnarray}}
\def\eeq{\end{equation}}
\def\eea{\end{eqnarray}}

\def\SM{Standard Model }
\title{\boldmath  Reduced modular symmetries of threshold corrections  and gauge coupling unification}
\author[a]{David Bailin}
\author[a]{Alex Love}
\affiliation[a]{Department of Physics \& Astronomy, \\ University of Sussex,\\ Brighton BN1 9QH, U.K.}
\emailAdd{d.bailin@sussex.ac.uk} 
\abstract{We revisit the question of gauge coupling unification at the string scale in orbifold compactifications of the heterotic string for the supersymmetric Standard Model. In the presence of discrete Wilson lines threshold corrections with modular symmetry that is a subgroup of the full modular group arise. We find that reduced modular symmetries not previously reported are possible. We conjecture that the effects of such threshold corrections can be simulated using sums of terms built from Dedekind eta functions to obtain the appropriate modular symmetry. For the cases of the $\mathbb{Z}_8$-I orbifold and the $\mathbb{Z}_3 \times \mathbb{Z}_6$ orbifold  it is easily possible to obtain gauge coupling unification at the ``observed'' scale with K\"ahler moduli $T$ of approximately one.}
\arxivnumber{}
\begin{document}
\maketitle
\flushbottom

%%%%%%%%%%%%%%%%%%%%

\section{Introduction}
It is well known that when the $SU(3)_{\rm colour} \times SU(2)_L \times U(1)_Y$ gauge coupling constants $g_{1,2,3}$ are extrapolated to high energies using the renormalisation group equations of the  supersymmetric \SM with just two Higgs doublets \cite{Ellis:1990wk,Amaldi:1991cn} the three gauge couplings reach a common value $g(m_X) =g_{\rm string}\simeq 0.7$ at a scale  $m_X \simeq 2 \times 10^{16}$ GeV. (Arguably, this is the best evidence to date for the presence of supersymmetry underlying the Standard Model.) In contrast, in the context of heterotic string theory, unification is expected \cite{Kaplunovsky:1987rp} at a scale $m_{\rm string} \simeq 5.27 \times 10^{17} g_{\rm string}$ \ GeV, a factor of order twenty or so larger than the ``observed'' unification scale. (Arguably, this is the best evidence to date that string theory is wrong!) However, the renormalisation group equations used in the extrapolation do not include contributions from the massive states associated with the unification threshold. In the case of string unification, since  infinite towers of massive states contribute above the string threshold, one might hope for substantial contributions $\Delta_a$  from such loops to the renormalisation group equations  for the gauge coupling constants $g_a(\mu), \ (a=1,2,3)$ of the (simple) group $G_a$
\beq
\frac{1}{g_a^2(\mu)}=\frac{k_a}{g_{\rm string}^2}+\frac{b_a}{16 \pi ^2}\ln \frac{m_{\rm string}^2}{\mu ^2}+\frac{1}{16 \pi ^2} \Delta_a \label{RGE}
\eeq
Here $k_a$ is the affine level of the gauge group factor $G_a$, 
\beq
b_a=-3C(G_a)+\sum _{{\bf R}_a}h_{{\bf R}_a}T({\bf R}_a)
\eeq
is the one-loop beta function coefficient for $G_a$, with $C(G_a)$ the quadratic Casimir for the adjoint representation, $T({\bf R}_a)={\rm Tr}  \ Q_a^2$ where $Q_a$ is any generator of $G_a$ in the representation ${\bf R}_a$, and $h_{{\bf R}_a}$ is the number of chiral multiplets in the representation $R_a$. W e use the standard normalisation of the generators such that for the fundamental representation ${\bf N}$ of the group $SU(N)$, $T({\bf N})=1/2$.

The threshold corrections $\Delta_a$ in general depend upon the various K\"ahler and complex structure moduli $T_i$ and $U_j$ associated with the 6-dimensional  space on which the heterotic string is compactified, and it is conceivable that these corrections can bridge the gap between the observed unification scale $m_X$ and the string scale $m_{\rm string}$, at least for certain, hopefully realistic, values of the moduli. 
Of course, this is not the only approach to solving this problem. As discussed in \cite{Dienes:1996du}, other possibilities include string GUT models, non-standard affine levels, light supersymmetry thresholds, extra non-supersymmetric matter, no-supersymmetric strings, and strong-coupling effects in string theory.

The toroidal world sheet associated with a (closed) string loop is characterised by a modular parameter $\tau$ and invariance under $PSL(2,\mathbb{Z})$  global modular transformations  of the form
\beq
\tau \rightarrow \frac{\alpha \tau-i\beta}{ i\gamma \tau + \delta} 
\eeq
with $ \alpha , \beta, \gamma , \delta \in \mathbb{Z}$ and $ \alpha   \gamma - \beta\delta =1$ in turn induces target-space invariance  under transformations of the moduli $T_k$ and $U_k$. Generically,  this modular symmetry is also $PSL(2,\mathbb{Z})$
\beq
T_k \rightarrow \frac{a_kT_k-ib_k}{ic_kT_k+d_k} \label{Tmod}
\eeq
with $a_k,b_k,c_k,d_k \in \mathbb{Z}$ and $a_kd_k-b_kc_k=1$.
Up to constant terms, independent of the moduli, the threshold correction arising from such an $\mathcal{N}=2$ sector has the form
\beq
\Delta_a=-({b'_a}^k-\delta_{\rm GS}^k) \ln\left[(T_k+\bar{T}_k)|\eta(T_k)|^4\right]
\eeq
with a similar contribution from the complex-structure modulus $U_k$. Here $\delta_{\rm GS}^k$ is the (universal) coefficient of the Green-Schwarz term, and  $\eta(T)$ is the Dedekind eta function
\beq
\eta(T):=e^{-\pi T/12}\prod_{n=1}^{\infty} (1-e^{-2 \pi nT})
\eeq
Note that for large values of $T$ the eta function $\eta(T) \sim e^{-\pi T/12}$, so that   $\Delta_a\propto T$. It is this feature that is exploited in attempts to bridge the gap between $m_X$ and $m_{\rm string}$. 
Under a modular transformation of the form given in eqn (\ref{Tmod}) 
\bea
&&T+\bar{T} \rightarrow \frac{ T+\bar{T}}{|icT+d|^2}\\
&&\eta(T) \rightarrow (icT+d)^{1/2}\eta(T)
\eea
up to a constant, so that $\Delta_a$ is indeed invariant.

 In this paper we are concerned only with orbifold compactifications in which the 6-torus $T^6$ is quotiented with a $\mathbb{Z}_N$ or $\mathbb{Z}_N \times \mathbb{Z}_M$ point group. For most of these there are $\theta^p$-twisted sectors  of the  generator(s) $\theta$ of the point group  that leave a single plane  invariant; such twisted sectors are referred to as $\mathcal{N}=2$ sectors.  The calculation of the contribution of a twisted sector with an invariant plane $T^2_k$  to the threshold corrections entails a sum over the momentum  $(m_{2k-1}, m_{2k})$ and winding numbers $(n_{2k-1},n_{2k})$ that arise in the zero-mode expansion  of the string world sheet in such sectors, and it is this that generates the dependence on the moduli $T_k$ and $U_k$. 
The foregoing result applies whenever  the $\mathcal{N}=2$ sector fixed planes are such that the torus $T^6$ on which the point group action is realised can be decomposed into a direct product $T^4 \oplus T^2$ with the fixed plane lying in $T^2$. This is certainly {\em not} the case for all orbifolds. Consider, for example,  the $\mathbb{Z}_6$-II orbifold. It has  point-group generator $\theta$ with eigenvalues $e^{2 \pi i v_k}, \ (k=1,2,3)$ where
\beq
(v_1,v_2,v_3)=\frac{1}{6}(2,1,-3)
\eeq
The symmetry may be realised on several lattices, but in particular on the $SU(3) \oplus G_2 \oplus SO(4)$  and $SU(6) \oplus SU(2)$  lattices; in both cases the action of the point-group is given by the Casimir operators of the various groups.  Evidently the $\theta^2, \theta ^4$ and $\theta^3$-twisted sectors all have fixed planes. In the first realisation, the lattice is the product of three 2-tori $T^2_1 \oplus T^2_2 \oplus T^2_3$, with $T^2_3$ the fixed plane of  $\theta^2, \theta ^4$, and $T^2_1$ the fixed plane of $\theta^3$. Since $\theta$ acts as $\mathbb{Z}_3$ on $T^2_1$, the  complex structure $U_1=e^{2\pi i/3}$ is fixed and there is no associated modular symmetry. On $T^2_3$, however, $\theta$ acts as a reflection $\mathbb{Z}_2$ and the complex structure $U_3$ is arbitrary.  In the first case, the above analysis applies, and there is the full $PSL(2,\mathbb{Z})$ modular symmetry for the K\"ahler moduli $T_1$ and $T_3$, as well as for the complex structure $U_3$. (Of course , there is also a K\"ahler modulus $T_2$ associated with the second plane, but since this is not an $\mathcal{N}=2$ plane, it does not enter the threshold corrections.) However, in the second case the lattice is $T^5\oplus T^1$ and the previous analysis does not apply. Such cases must be treated on a case by case basis, and in this case it was shown \cite{Bailin:1993wv} that the modular group associated with $T_1$ is $\Gamma^0(2)$, meaning the subgroup of $PSL(2,\mathbb{Z})$ defined in eqn (\ref{Tmod}) with the integer $b_1=0 \bmod 2$. Similarly, the modular group associated with $T_3$ is $\Gamma^0(3)$, corresponding to the subgroup with $b_3=0 \bmod 3$. 
The modular symmetry associated with  $U_3$ was shown to be $\Gamma_0(3)$, corresponding to $c'_3=0 \bmod 3$, with $c'_3$ the analogue for $U_3$ of $c_3$.  As in the separable case, the threshold corrections $\Delta _a$ may be calculated and are indeed invariant under the reduced modular symmetries. For example, the $T_3$- and $U_3$-dependent threshold corrections are \cite{Bailin:1993fm}
\bea
\Delta_a=&&-2b_a^{(1,\theta^2)}\ln \left[(T_3+\bar{T}_3)|\eta(T_3)|^4(U_3+\bar{U}_3)|\eta(U_3)|^4\right] \nonumber \\
&&-2b_a^{(1,\theta^2)}\ln \left[(T_3+\bar{T}_3)\left|\eta\left(\frac{T_3}{3}\right)\right|^4(U_3+\bar{U}_3)|\eta(3U_3)|^4\right]  \label{Dela}
\eea

The $PSL(2,\mathbb{Z})$ modular symmetries are also broken by the presence of discrete background Wilson lines \cite{Spalinski:1991vd,Erler:1991ju,Ibanez:1992hc, Erler:1992av,Bailin:1993ri}. Besides the standard degrees of freedom $X^i, \Psi^i$  associated with the 10-dimensional string, the heterotic string has 16 additional left-moving bosonic degrees of freedom  $X^{I}(\tau+ \sigma)$ on the $E_8 \times E_8$ lattice. Background Wilson lines $A^I_a$  couple the $X^I$ to the six bosonic degrees of freedom $X^a$ associated with the  (orbifold) compactified dimensions via a term in the world-sheet action
\beq
S \supset \frac{1}{2\pi} \int d\tau d\sigma \epsilon^{\alpha \beta}A^I_a \partial _{\alpha} X^a \partial _{\beta} X^I
\eeq 
This has the effect of shifting the canonical left- and right-moving momenta by terms involving the quantised momentum on the $E_8 \times E_8$ lattice, but also of shifting the lattice momentum by terms involving the winding numbers. It is this mixing that modifies the modular-invariance properties. However, unlike the situation in the absence of Wilson lines, to our knowledge the string loop threshold corrections in the presence of discrete background Wilson lines  have not actually been calculated\footnote{The matter was   addressed by Kaput \& Paleani in 2010 \cite{Klaput:2010dg}, but we are unclear as to the status of this paper since it has not been published. We are grateful to the referee for drawing our atention to it.}. 
A systematic study of the modular symmetries in the presence of discrete Wilson lines was made in \cite{Love:1996sk} for all Abelian orbifolds whose point-group action is realised by Coxeter elements or their generalisations.  

In this paper we shall revisit this question, but only for those orbifolds in which the $\mathcal{N}=2$  fixed planes are separable into the $T^2$ part of the 6-torus $T^4 \oplus T^2$ on which the point group is realised. We  also restrict ourselves to models with {\em no} $\mathbb{Z}_2$  fixed planes; as is apparent from the previous treatment \cite{Love:1996sk}, such models are considerably more complicated. In the notation of that paper, only (certain compactifications of) the $\mathbb{Z}_6$-I, $\mathbb{Z}_8$-I, $\mathbb{Z}_{12}$-I,  $\mathbb{Z}_3 \times \mathbb{Z}_3$, $\mathbb{Z}_4 \times \mathbb{Z}_4$, $\mathbb{Z}_3 \times \mathbb{Z}_6$ and $\mathbb{Z}_6 \times \mathbb{Z}_6$ orbifolds satisfy these criteria.
We also
speculate on the effect that our new results might have on the question of whether  the threshold corrections can bridge the gap between the observed unification scale $m_X$  of the supersymmetric \SM and $m_{\rm string}$. In the absence of discrete Wilson lines, such unification can only be achieved with large values $T \sim 26$ of the K\"ahler moduli \cite{Ibanez:1992hc}. In any case, it was also shown that this unification is not possible for the $\mathbb{Z}_6$-I or $\mathbb{Z}_{12}$-I orbifolds.  Further, for the  $\mathbb{Z}_N \times \mathbb{Z}_M$ orbifolds, whose threshold corrections depend upon all three K\"ahler moduli $T_{1,2,3}$, it is reasonable, in our view, to insist that unification is achievable with all three moduli of comparable size $T_1\simeq T_2\simeq T_3$, in which case of these only the $\mathbb{Z}_3 \times \mathbb{Z}_6$ survives. Of course, one may take the opposite view to ours and make a virtue of highly anisotropic compactifications; see, for  example, Hebecker \& Trapletti \cite{Hebecker:2004ce}.

Thus, in the end, we pursue the question of threshold corrections just for the $\mathbb{Z}_8$-I and $\mathbb{Z}_3 \times \mathbb{Z}_6$ orbifolds. We shall argue that the considerably reduced modular symmetry that we now find can arise when discrete Wilson lines are present  allows unification in both models with K\"ahler moduli $T \sim 1$ of order unity, if we conjecture that the effect of such threshold corrections can be simulated using sums of terms built from Dedekind eta functions. There have been some attempts to construct realistic models on both of these orbifolds; see, for example, Groot Nibbelink \& Loukas \cite{Nibbelink:2013lua} on the $\mathbb{Z}_8$ orbifold and Niles \& Vaudrevange \cite{Nilles:2014owa} on the $\mathbb{Z}_N \times \mathbb{Z}_M$ orbifolds, and it will be interesting to see whether our results apply to their models.
%%%%%%%%%%%%%%%%%%%%%%%%%%%%%%%%%%%%%%%%%%%%%%%%%%%%%%%%%%%%%%%%%%%%%%%%%%%%%%%%%%%%%%%%%%%%%%%%%%%%%%%%%%%%%%%%%%%%%%%%%%%%%%%%%%%%%%
\section{Duality symmetry and Wilson lines}  
The action of the point group generator $\theta$  of a $\mathbb{Z}_N$ orbifold on the defining lattice basis ${\bf e}_a, \ (a=1,2, ...\ 6)$, is given by the elements of a matrix $Q$ defined by
\beq
\theta {\bf e}_a=Q_{ba}{\bf e}_b
\eeq
The $16\times 6$ dimensional Wilson line matrix $A$ with elements $A^I_a$ may be written as
\beq
A:= ({\bf A}_1,{\bf A}_2,{\bf A}_3,{\bf A}_4,{\bf A}_5,{\bf A}_6)
\eeq                 
where ${\bf A}_a$ are $16$-dimensional column vectors. It is required to satisfy \cite{Erler:1992av}
\beq
A(1-Q) \in {\bf \mathbb{Z}} \label{AQ}
\eeq
and this leads to the constraints on the vectors ${\bf A}_a$ that are listed in Appendix B of \cite{Love:1996sk}. $A$ must also satisfy
\beq
A^tCA(1-Q)+(1-Q^*)A^tCA \in {2\mathbb{Z}} \label{QQst}
\eeq
where $C$ is the Cartan matrix for $E_8 \times E_8$ and 
\beq
Q^*:=(Q^t)^{-1}
\eeq
For the $\mathbb{Z}_N \times \mathbb{Z}_M$ orbifolds, analogous  constraints apply to the matrices $Q_{\theta}$ and $Q_{\omega}$ representing the two generators $\theta$ and $\omega$. 

As detailed in Appendix B of \cite{Love:1996sk}, the fixed planes of the various orbifolds may be  classified according to how the point-group generator(s) acts on that plane. The only possibilities are $\mathbb{Z}_2,\ \mathbb{Z}_3,\ \mathbb{Z}_4$ and $ \mathbb{Z}_6$. How they arise in the $T^4 \oplus T^2$ separable orbifolds is indicated in Table \ref{ZN}. Also displayed are the restrictions following from eqn (\ref{AQ}) on the Wilson lines $({\bf A}_{2k-1}, {\bf A}_{2k})$ in that plane. (The twists associated with the point-group generators $\theta$ (and $\omega$) are listed in Appendix A of \cite{Love:1996sk}, for example.)
\begin{table}
 \begin{center}
\begin{tabular}{||c|c|c|c||} \hline \hline
$\mathbb{Z}_N$ fixed plane & Orbifold &Twisted sector& $({\bf A}_{2k-1}, {\bf A}_{2k})$ \\ \hline \hline 
%%%%%%%%%%%%%%%%%%%%%%%%%%%%%%%%%%%%%%%%%%%%%%%%%%%%%%%%%%%%%%%%%%%%%%%%%%%%%%%%%%%%%%%%%%%%%%%%%%%%%%%%%%%%%%
$\mathbb{Z}_2$&$\mathbb{Z}_4$& $\theta ^2$ & $2{\bf A}_3=0 \bmod 1=2{\bf A}_6$ \\
&$\mathbb{Z}_6$-II & $\theta ^2$&  $2{\bf A}_5=0 \bmod 1=2{\bf A}_6$ \\
&$\mathbb{Z}_8$-II & $\theta ^2$&  $2{\bf A}_5=0 \bmod 1=2{\bf A}_6$ \\
&$\mathbb{Z}_{12}$-II & $\theta ^2$&  $2{\bf A}_5=0 \bmod 1=2{\bf A}_6$ \\
\hline
$\mathbb{Z}_3$ & $\mathbb{Z}_6$-I,II & $\theta ^3$&   ${\bf A}_1-{\bf A}_2=0 \bmod 1, 3{\bf A}_1=0 \bmod 1$ \\ 
&$\mathbb{Z}_{12}$-I& $\theta ^3$& ${\bf A}_5-{\bf A}_6=0 \bmod 1, 3{\bf A}_5=0 \bmod 1$ \\
&$\mathbb{Z}_3 \times \mathbb{Z}_3$&$\omega^k$&${\bf A}_1-{\bf A}_2=0 \bmod 1, 3{\bf A}_1=0 \bmod 1$ \\
&&$\theta^k$&${\bf A}_3-{\bf A}_4=0 \bmod 1, 3{\bf A}_3=0 \bmod 1$ \\
&&$(\theta\omega^2)^k$&${\bf A}_5-{\bf A}_6=0 \bmod 1, 3{\bf A}_5=0 \bmod 1$ \\
&$\mathbb{Z}_3 \times \mathbb{Z}_6$& $\omega^k$& ${\bf A}_1-{\bf A}_2=0 \bmod 1, 3{\bf A}_1=0 \bmod 1$ \\ \hline
$\mathbb{Z}_4$ &$\mathbb{Z}_8$-I &$\theta^4$&${\bf A}_5=0 \bmod 1= 2{\bf A}_6$ \\
&$\mathbb{Z}_4 \times \mathbb{Z}_4$&$\omega^k$ & ${\bf A}_1=0 \bmod 1= 2{\bf A}_2$ \\
&&$\theta^k$ & ${\bf A}_3=0 \bmod 1= 2{\bf A}_4$ \\
&&$(\theta^2\omega^2)^k$ & ${\bf A}_5=0 \bmod 1= 2{\bf A}_6$ \\ \hline
$\mathbb{Z}_6$ &$\mathbb{Z}_3 \times \mathbb{Z}_6$&$\theta^k$ & ${\bf A}_3=0 \bmod 1= {\bf A}_4$ \\
&&$(\theta\omega^4)^k$ & ${\bf A}_5=0 \bmod 1= {\bf A}_6$ \\ 
&$\mathbb{Z}_6 \times \mathbb{Z}_6$&$\omega^k$ & ${\bf A}_1=0 \bmod 1= {\bf A}_2$ \\
&&$\theta^k$ & ${\bf A}_3=0 \bmod 1= {\bf A}_4$ \\
&&$(\theta\omega^5)^k$ & ${\bf A}_5=0 \bmod 1= {\bf A}_6$ \\
\hline \hline
\end{tabular}
\end{center} 
\caption{ \label{ZN}  Classification of fixed planes by the action of the point group on that plane.}
 \end{table}
%%%%%%%%%%%%%%%%%%%%%%%%%%%%%%%%%%%%%%%%%%%%%%%%%%%%%%%%%%%%%%%%%%%%%%%%%%%%%%%%%%%%%%%%%%%%%%%%%%%%%%%%%%%%%%%%

The modular symmetry of the the K\"ahler moduli $T_k$ associated with the fixed torus $T^2_k$, in which $T_k$ transforms as in eqn (\ref{Tmod}), 
is constrained \cite{Bailin:1993ri} by
\bea
&&c_k\left({\bf A}_{2k-1},{\bf A}_{2k} \right) \in \mathbb{Z} \label{cA} \\
&&c_k\left( \begin{array}{c} {\bf A}_{2k-1}^t \\
 {\bf A}_{2k}^t
\end{array} \right)C\left({\bf A}_{2k-1},{\bf A}_{2k}\right) \in 2\mathbb{Z} \label{ACA} \\
&&c_k\left({\bf A}_{2k-1},{\bf A}_{2k}\right)J\left( \begin{array}{c} {\bf A}_{2k-1}^t \\
 {\bf A}_{2k}^t
\end{array} \right)C \in \mathbb{Z} \label{AJA}\\
&&2(1-d_k)\left({\bf A}_{2k-1},{\bf A}_{2k} \right)-c_k\left({\bf A}_{2k-1},{\bf A}_{2k} \right)J\left( \begin{array}{c} {\bf A}_{2k-1}^t \\
 {\bf A}_{2k}^t
\end{array} \right)C\left({\bf A}_{2k-1},{\bf A}_{2k} \right)\in 2\mathbb{Z}   \label{dc}\\
&&2(1-a_k)C\left({\bf A}_{2k-1},{\bf A}_{2k} \right)-c_kC\left({\bf A}_{2k-1},{\bf A}_{2k} \right)J\left( \begin{array}{c} {\bf A}_{2k-1}^t \\
{\bf A}_{2k}^t
\end{array} \right)C\left({\bf A}_{2k-1},{\bf A}_{2k} \right)\in 2\mathbb{Z}\label{ac} \nonumber \\
&& \\
&&2(2-a_k-d_k)\left( \begin{array}{c} {\bf A}_{2k-1}^t \\
 {\bf A}_{2k}^t
\end{array} \right)C\left({\bf A}_{2k-1},{\bf A}_{2k} \right) \nonumber\\
&& \qquad \qquad-c_k\left( \begin{array}{c} {\bf A}_{2k-1}^t \\
 {\bf A}_{2k}^t
\end{array} \right)C\left({\bf A}_{2k-1},{\bf A}_{2k} \right)J\left( \begin{array}{c} {\bf A}_{2k-1}^t \\
 {\bf A}_{2k}^t
\end{array} \right)C\left({\bf A}_{2k-1},{\bf A}_{2k} \right)\in 4\mathbb{Z} \label{adc}
\eea
where
\beq
J:=\left(\begin{array}{cc} 0&1 \\
-1&0 \end{array} \right)
\eeq

%%%%%%%%%%%%%%%%%%%%%%%%%%%%%%%%%%%%%%%%%%%%%%%%%%%%%%%%
We shall discuss in turn the modular symmetries of  the three non-$\mathbb{Z}_2$ fixed plane classes displayed in Table \ref{ZN}.
\subsection{ $\mathbb{Z}_6$-planes}
 Consider first the $\mathbb{Z}_6$-plane $T^2_2$ arising in the $\theta^k$-twisted sectors of the $\mathbb{Z}_3 \times \mathbb{Z}_6$ orbifold. Eqn (\ref{AQ}) requires only that the Wilson lines in that torus are $E_8 \times E_8$ lattice vectors. They are also constrained by eqn (\ref{QQst}) which in this case gives
\beq
\left( \begin{array}{cc} {\bf A}_3^tC{\bf A}_3&-{\bf A}_3^tC{\bf A}_3-{\bf A}_4^tC{\bf A}_4 \\
 {\bf A}_3^tC{\bf A}_3+2{\bf A}_3^tC{\bf A}_4 +{\bf A}_4^tC{\bf A}_4 &{\bf A}_4^tC{\bf A}_4 
\end{array} \right)\in 2\mathbb{Z}  \label{QQst34}
\eeq
which is automatically satisfied when ${\bf A}_3$ and ${\bf A}_4$ are lattice vectors. Eqn (\ref{cA}) is also satisfied since $c_2$ is an integer. However, a new constraint arises from eqn (\ref{ACA}), namely that 
\beq
c_2 {\bf A}_3^tC{\bf A}_4 \in 2\mathbb{Z}
\eeq
Since ${\bf A}_3$ and ${\bf A}_4$ are lattice vectors, ${\bf A}_3^tC{\bf A}_4 \in \mathbb{Z}$, so if ${\bf A}_3^tC{\bf A}_4$ is even, then eqn (\ref{AJA}) is satisfied without any further constraint on $c_2$. Alternatively,  if ${\bf A}_3^tC{\bf A}_4$ is odd, then we require that $c_2 = 0 \bmod 2$. In both cases, eqns 
(\ref{AJA}) ... (\ref{ac}) are then satisfied without further constraint. If ${\bf A}_3^tC{\bf A}_4$ is even, eqn (\ref{adc}) is also satisfied without further constraint. When ${\bf A}_3^tC{\bf A}_4$ is odd, then since $c_2$ must be even it follows from the unimodularity that $a_2$ and $d_2$ are both odd, and this ensures that the first term on the left-hand side of eqn (\ref{adc}) is $0 \bmod 4$. However, ensuring that the second term is also $0 \bmod 4$  requires that $c_2=0 \bmod 4$. Thus there are two generic cases for the modular symmetry of $T_2$. If   ${\bf A}_3^tC{\bf A}_4$ is even, then it is the full $PSL(2,\mathbb{Z})$. Otherwise it is $\Gamma_0(4)$, the subgroup of $PSL(2,\mathbb{Z})$ with $c_2=0 \bmod 4$. This alternative  was overlooked in reference \cite{Love:1996sk}.

The torus $T^2_3$ is fixed in the $(\theta \omega^4)^k$-twisted sectors. Since the action of $\theta$ in $T^2_3$ is just the square of the action of $\omega$, this too is a $\mathbb{Z}_6$-plane. The action of $\omega$ in $T^2_3$ is the same as its action in $T^2_2$, so the constraint deriving from eqn (\ref{QQst}) is the same as that displayed in eqn (\ref{QQst34}) but with $({\bf A}_5, {\bf A}_6)$ replacing $({\bf A}_3, {\bf A}_4)$. It follows again that there are   two generic cases for the modular symmetry of $T_3$. If   ${\bf A}_5^tC{\bf A}_6$ is even, then it is the full $PSL(2,\mathbb{Z})$. Otherwise it is $\Gamma_0(4)$. The same analysis applies to the three  $\mathbb{Z}_6$-planes of the $\mathbb{Z}_6\times \mathbb{Z}_6$ orbifold, so the conclusion is that these are the two generic symmetries for {\em all} (K\"ahler) moduli in $\mathbb{Z}_6$-planes.
%%%%%%%%%%%%%%%%%%%%%%%%%%%%
\subsection{$\mathbb{Z}_4$-planes}
For the fixed plane $T^2_3$ that arises in the $\theta^4$-twisted sector of the $\mathbb{Z}_8$ orbifold, the first $\mathbb{Z}_4$-plane entry in Table \ref{ZN}, the constraint given in  eqn (\ref{QQst}) yields

\beq
\left( \begin{array}{cc} 2{\bf A}_{5}^tC{\bf A}_{5} & {\bf A}_{5}^tC{\bf A}_{5} +4{\bf A}_{5}^tC{\bf A}_{6} +2{\bf A}_{6}^tC{\bf A}_{6 } \\
-{\bf A}_{5}^tC{\bf A}_{5} -2{\bf A}_{6}^tC{\bf A}_{6}  & 2{\bf A}_{6}^tC{\bf A}_{6} 
\end{array}  \right) \in 2\mathbb{Z}
\eeq
where ${\bf A}_{5}$ and $2{\bf A}_{6}$ are lattice vectors.  This is satisfied provided only that
\beq
 {\bf A}_{6}^tC{\bf A}_{6} \in \mathbb{Z} \label{A66}
\eeq
If ${\bf A}_{6}$ is a lattice vector, then eqn (\ref{A66}) is automatically satisfied, and the analysis of the modular constraints is the same as that given above for the $\mathbb{Z}_6$-planes. In this case, then, the conclusions are identical.

We therefore need only consider the alternative, namely that ${\bf A}_{6}$ is {\em not} a lattice vector. 
  In this case, eqn (\ref{cA}) requires that $c_3=0 \bmod 2$. Then,  the only further constraint from eqn (\ref{ACA}) is that $c_3{\bf A}_{5}^tC{\bf A}_{6}$ is even. Now, ${\bf A}_{5}^tC{\bf A}_{6}$ is either an integer or half-integer. If the former,  then there is no further constraint on $c_3$. However, if  ${\bf A}_{5}^tC{\bf A}_{6}\in \mathbb{Z}+\frac{1}{2}$, then we require that $c_3=0 \bmod 4$. In both cases, eqn (\ref{AJA}) is satisfied without further constraint. Since $c_3$ is even, unimodularity requires that $a_3$ and $d_3$ are both odd, and the first term on the left-hand side of eqns (\ref{dc}) and (\ref{ac}) are both even. Then, to satisfy eqn (\ref{dc}) we only require that $c_3{\bf A}_{5}^tC{\bf A}_{6}=0\bmod4$. If ${\bf A}_{5}^tC{\bf A}_{6}$ is even, this is already satisfied. However, if ${\bf A}_{5}^tC{\bf A}_{6}$ is odd, we now require that $c_3=0 \bmod 4$, whereas if  ${\bf A}_{5}^tC{\bf A}_{6}\in \mathbb{Z}+\frac{1}{2}$, then we require that $c_3=0 \bmod 8$. These are sufficient to ensure that  eqn (\ref{ac}) is fully satisfied and, except in the third eventuality, that eqn (\ref{adc}) is too. In the case that  ${\bf A}_{5}^tC{\bf A}_{6}\in \mathbb{Z}+\frac{1}{2}$, this requires that $c_3= 0 \bmod 16$. Thus, when ${\bf A}_{6}$ is {\em not} a lattice vector  the conclusions are that (i) if ${\bf A}_{5}^tC{\bf A}_{6}$ is even, the modular group for $T_3$ is $\Gamma_0(2)$, (ii) if ${\bf A}_{5}^tC{\bf A}_{6}$ is odd, then the modular group is $\Gamma_0(4)$, but  (iii) if  ${\bf A}_{5}^tC{\bf A}_{6}\in \mathbb{Z}+\frac{1}{2}$, then it is $\Gamma_0(16)$.
%%%%%%%%%%% 
%need to cover the other Z4-plane cases too
%%%%%%%%%%%%%
\subsection{$\mathbb{Z}_3$-planes}
For the fixed plane $T^2_1$ that arises in the $\omega^k$-twisted sector of the $\mathbb{Z}_3 \times \mathbb{Z}_6$ orbifold, the last $\mathbb{Z}_3$-plane entry in Table \ref{ZN}, the constraint given in  eqn (\ref{QQst}) yields
\beq
\left( \begin{array}{cc} 3{\bf A}_1^t C{\bf A}_1 &{\bf A}_1^t C{\bf A}_1+4{\bf A}_1^t C{\bf A}_2+{\bf A}_2^t C{\bf A}_2 \\
 2{\bf A}_1^t C{\bf A}_2-{\bf A}_1^t C{\bf A}_1-{\bf A}_2^t C{\bf A}_2&3{\bf A}_2^t C{\bf A}_2 
\end{array} \right)\in 2\mathbb{Z} \label{T21AQ}
\eeq
where ${\bf A}_1$ and ${\bf A}_2$  differ by a lattice vector and $3{\bf A}_1$ is a lattice vector. As above, if ${\bf A}_{1}$,  and hence ${\bf A}_{2}$, are  lattice vectors, then this   is automatically satisfied, and the analysis of the modular constraints is the same as for the $\mathbb{Z}_4$-planes. The conclusions, in this case for $T_1$, are the same.

We therefore need only consider the alternative, namely that ${\bf A}_{1}$ is {\em not} a lattice vector. Then to satisfy eqn (\ref{T21AQ}) we must have that
\beq
3{\bf A}_1^t C{\bf A}_1 \in 2\mathbb{Z}
\eeq
We may write 
\beq
{\bf A}_2-{\bf A}_1:={\bf V}
\eeq
where ${\bf V} \in \mathbb{Z}$ is a lattice vector, so that ${\bf V}^tC{\bf V} \in 2 \mathbb{Z}$;
it follows that $3{\bf A}_2^t C{\bf A}_2 $ is also even, since $3{\bf A}_1^t C{\bf V} $ is an integer. Eqn (\ref{cA}) requires that $c_1=0 \bmod 3$. Then,  the only further constraint from eqn (\ref{ACA}) is that $c_1{\bf A}_{1}^tC{\bf V}=0 \bmod 2$. So, if   $3{\bf A}_{1}^tC{\bf V}=3{\bf A}_1^t C{\bf A}_2 \bmod 2$ is even, there is no further constraint on $c_1$, but if $3{\bf A}_{1}^tC{\bf V}$ is odd, then we require that $c_1$ is even, and hence $c_1=0 \bmod 6$. As before, eqn (\ref{AJA}) is then satisfied  without further constraint. In this case, unimodularity now means that we may choose that $a_1= 1 \bmod 3=d_1$, which ensures that the first term on the left-hand sides of eqns (\ref{dc}) and (\ref{ac}) are even. Then to satisfy them we require only that 
$c_1{\bf A}^t_1C{\bf V}=0 \bmod 6$. Thus, if $3{\bf A}_{1}^tC{\bf V}=0\bmod 6$ there is no further constraint; if $3{\bf A}_{1}^tC{\bf V}=0\bmod 2$ (but not $0\bmod 6$), then $c_1=0 \bmod 9$. However  if $3{\bf A}_{1}^tC{\bf V}$ is odd, we require that $c_1=0 \bmod 18$. 
 Finally, unimodularity requires that $2-a_1-d_1=0 \bmod c_1$, and to satisfy eqn (\ref{adc}) we require only that $c_1({\bf A}^t_1C{\bf V})^2=0 \bmod 4$. In the case that $3{\bf A}_{1}^tC{\bf V}$ is even, there is no further constraint on $c_1=0 \bmod 9$. However, if $3{\bf A}_{1}^tC{\bf V}$ is odd, we now require that $c_1=0 \bmod 36$.  In the former case, the modular symmetry for $T_1$ is $\Gamma_0(9)$, and in the latter it is $\Gamma_0(36)$.

It is the availability of these large integers $n$ in the reduced modular symmetry $\Gamma _0(n)$ of the $\mathbb{Z}_8$ and $\mathbb{Z}_3 \times \mathbb{Z}_6$ orbifolds that we believe may enable the gap between $m_X$ and $m_{\rm string}$ to be bridged.
%%%%%%%%%%%%%%%%%%%%%%%%%%%%%%%%%%%%%%%%%%%%%%%%%%%%%%%%%%%%%%%%%%%%%%%%%%%%%%%%%%%%%%%%%%%%%%%%%%%%%%%%%%%%%%%%%%%%%%%%%%%%%%%%%%%%
\section{Gauge coupling constant  unification}
The question we shall now address is whether the inclusion of the moduli-dependent threshold corrections in the renormalisation group equations (\ref{RGE}) allows  unification of the  \SM gauge coupling constants $g_a(\mu), \ (a=1,2,3)$  at a scale $m_X \simeq 2 \times 10^{16}$ GeV consistently with the string scale $m_{\rm string} \simeq 3.7\times 10^{17}$ GeV. Taking all groups at level 1, this requires that for $a,b=1,2,3$
\beq
 \frac{m_{\rm string}^2}{m_X^2}=\exp \left(-\frac{\Delta_a- \Delta_b}{b_a-b_b}\right) \label{ms2mx2}
\eeq
and consistency requires that 
\beq
\frac{\Delta _3 - \Delta_2}{\Delta _3 - \Delta_1}=\frac{b_3-b_2}{b_3-b_1}= \frac{5}{12} \label{Del321}
\eeq
in the \SM since
\beq
b_3=-3, \ b_2=1, \ b_1 = \frac{33}{5}
\eeq
Since $m_{\rm string}>m_X$, we also require that 
\beq
\Delta _3 - \Delta_2>0 \label{Del32}
\eeq 
so that the exponent in eqn (\ref{ms2mx2}) is positive.

As we have already noted, the precise form of the threshold corrections in the presence of discrete Wilson lines is not known. However, since we know their modular symmetry, we may make plausible hypotheses of their form. In this paper we are only concerned with orbifolds in which there are no $\mathbb{Z}_2$ planes, in which case the complex structure moduli are fixed and only the modular symmetry of the K\"ahler moduli is relevant. In  analogy with eqn (\ref{Dela}) we assume\footnote{The form assumed appears to be consistent with the results of \cite{Klaput:2010dg}.} that $\Delta _a$ has the general form
\beq
\Delta_a= -\sum_{k=1}^3 ({b'_a}^k-\delta_{\rm GS}^k) \ln [(T_k+\bar{T}_k)|\eta(p_kT_k)\eta(q_kT_k)|^2]
\eeq
where $p_k$ and $q_k$ are integers consistent with the modular symmetry in the $\mathcal{N}=2$ plane $T^2_k$. The coefficients ${b'_a}^k$ are given by the general formula \cite{Ibanez:1992hc}
\beq
{b_a'}^k=-c(G_a)+ \sum _{{\bf R}_a}T({\bf R}_a)(1+2n^k_{{\bf R}_a})
\eeq
%Then eqn (\ref{ms2mx2})  may be written as
%\beq
%2\ln\frac{m_{\rm string}}{m_X}=   \sum_{k=1}^3 \frac{{b'_a}^k-{b'_b}^k}{b_a-b_b} \ln [(T_k+\bar{T}_k)|\eta(a_kT_k)\eta(b_kT_k)|^2]
%\eeq
Thus,
\bea
&&{b_3'}^k=3+\sum_{g=1}^3(2n^k_{Q_g}+n^k_{u_g}+n^k_{d_g}) \label{b3}\\
&&{b_2'}^k=5+n_h^k+n_{\bar{h}}^k+\sum_{g=1}^3(3n^k_{Q_g}+n^k_{L_g})\\
&&{b_1'}^k=\frac{33}{5}+\frac{3}{5}(n_h^k+n_{\bar{h}}^k)+\frac{1}{5}\sum_{g=1}^3(n^k_{Q_g}+8n^k_{u_g}+2n^k_{d_g}+3n^k_{L_g}+6n^k_{\ell_g}) \label{b1}
\eea
where $g=1,2,3$ labels the three \SM fermion generations, with $Q_g$ and $L_g$ respectively the quark  and lepton doublets,  $u_g,d_g$ and $\ell_g$ the quark and lepton singlets, and $h,\bar{h}$ are the two Higgs doublets;
$n^k_X$ is the modular weight of the state $X$ under the modular transformation given in eqn (\ref{Tmod}).
%%%%%%%%%%%%%%%%%%%%%%%%%%%%%%%%%%%%%%%%%%%%%%
\subsection{The $\mathbb{Z}_8$-I orbifold}
The point group generator $\theta$ has eigenvalues $e^{2\pi i v_k}$ with
\beq
(v_1,v_2,v_3)= \frac{1}{8}(1,-3,2)
\eeq
This may be realised by Casimirs acting  on the $SO(9) \times SO(5)$ lattice or on the $SO(8)^{[2]} \times SO(5)$ lattice\footnote{These two lattices are actually equivalent, as is shown in Table 5.4 of reference Fischer {\it et al.} \cite{Fischer:2012qj}. We are grateful to the referee for drawing our attention to this paper.
}, both of which have the form $T^4 \times T^2$. In both there is a single fixed $\mathcal{N}=2$ plane $T^2_3$ in the $\theta^4$-twisted sector. 
 Since $\theta$ acts as $\mathbb{Z}_4$ on this plane, only the K\"ahler modulus $T_3$ enters the threshold corrections. In this case eqns (\ref{Del32})  and  (\ref{Del321}) require that 
\bea
{b_3'}^3-{b_2'}^3>0 \label{b32}\\
{b_3'}^3-{b_1'}^3>0 \label{b31}\\
\frac{{b_3'}^3-{b_2'}^3}{{b_3'}^3-{b_1'}^3}=\frac{5}{12} \label{512}
\eea
and eqns (\ref{b3}), .. (\ref{b1}) give
\bea
&&{b_3'}^3-{b_2'}^3=-2-Y^3+\sum_{g=1}^3(-n^3_{Q_g}+n^3_{u_g})>0  \label{b320}\\
&& \frac{5}{3}({b_3'}^3-{b_1'}^3) =    -6-Y^3+\sum_{g=1}^3(3n^3_{Q_g}-n^3_{u_g}-2n^3_{\ell_g})>0 \label{b310}
\eea
where
\beq
Y^k:= n_h^k+n_{\bar{h}}^k-\sum_{g=1}^3(n^k_{d_g}-n^k_{L_g})  \quad (k=1,2,3) \label{Xi}
\eeq
Then, eliminating $Y^3$ (but not yet imposing the constraint in eqn (\ref{512})) gives
\beq
\frac{5}{3}({b'}^3_{3}- {b'}^3_{1})-({b'}^3_{3}- {b'}^3_{2})=-4+2\sum_{g=1}^3(2n^3_{Q_g}-n^3_{u_g}-n^3_{\ell_g}) \label{sumgn35a} 
\eeq
The calculation of the modular weights of the various states on this orbifold was treated in references \cite{Ibanez:1992hc,Bailin:1992hx} and they are displayed in Table \ref{n123}. It is easy to satisfy these constraints, including eqn (\ref{512}), for several values of 
\beq
{b'}^3_{3}- {b'}^3_{2}=\frac{1}{2}(1,2,3,4) \label{b321234}
\eeq
%%%%%%%%%%%%%%%%%%%%%%%%%%%%%%%%%%%%%%%
\begin{table}
 \begin{center}
\begin{tabular}{||c||c|c|c|c||} \hline \hline
State $X$ & $-n_X^1$ & $-n_X^2$ & $-n_X^3$ &$-\sum_k n^k_X$ \\ \hline \hline
 $Q, u, \ell$ \ {\rm and} \ $L,h,\bar{h}, d$&1&0&0 &1\\
&0&1&0&1 \\
&0&0&1&1 \\
& $\frac{1}{2}$ & $\frac{1}{2}$ & 0&1 \\ \hline
%%%%%%%%%
& $\frac{7}{8}$ & $\frac{3}{8}$ & $\frac{3}{4}$ &2\\
& $\frac{3}{8}$ & $\frac{7}{8}$ & $\frac{3}{4}$&2 \\
& $\frac{3}{4}$ & $\frac{3}{4}$ & $\frac{1}{2}$&2 \\ \hline
%%%%%%%%
& $\frac{15}{8}$ & $\frac{3}{8}$ & $\frac{3}{4}$&3 \\
& $\frac{3}{8}$ & $\frac{15}{8}$ & $\frac{3}{4}$&3 \\ 
 \hline \hline
%%%%%%
{\rm Only} \ $L,h,\bar{h}, d$ & $\frac{7}{4}$ & $\frac{3}{4}$ & $\frac{1}{2}$&3 \\
& $\frac{3}{4}$ & $\frac{7}{4}$ & $\frac{1}{2}$&3 \\
& $\frac{3}{8}$ & $\frac{7}{8}$ & $\frac{7}{4}$&3 \\
&$\frac{7}{8}$ & $\frac{3}{8}$ & $\frac{7}{4}$&3 \\ \hline
&$\frac{23}{8}$ & $\frac{3}{8}$ & $\frac{3}{4}$&4 \\
& $\frac{3}{8}$ & $\frac{23}{8}$ & $\frac{3}{4}$&4 \\
\hline \hline
\end{tabular}
\end{center} 
\caption{ \label{n123}    Modular weights of matter representations on the $\mathbb{Z}_8$ orbifold.}
 \end{table}
%%%%%%%%%%%%%%%%%%%%%%%%%%%%%%%%%%%%%%%%%%%%%%%%%%%%

The cancellation of target-space anomalies in $T^4$ requires also that ${b'}^k_{a}=\delta_{\rm GS}^k$ independently of the gauge group $a$. Hence
\beq
{b'}^k_{3}={b'}^k_{2}={b'}^k_{1} \quad (k=1,2) \label{modanom}
\eeq
and 
\beq
{b'}^k_{3}-{b'}^k_{2}=0={b'}^k_{3}-{b'}^k_{1} \quad (k=1,2)
\eeq
Then summing over $k$ gives
\bea
&&{b_3'}^3-{b_2'}^3=-6-\sum_{k=1}^3Y^k+\sum_{g,k=1}^3(-n^k_{Q_g}+n^k_{u_g})>0  \label{b32y}\\
&& \frac{5}{3}({b_3'}^3-{b_1'}^3) =    -18-\sum_{k=1}^3Y^k+\sum_{g,k=1}^3(3n^k_{Q_g}-n^k_{u_g}-2n^k_{\ell_g})>0 \label{b31y}
\eea
Now, it is clear from Table \ref{n123} that the sum $\sum _kn^k_X $ of the modular weights for any particle is an integer, and in fact this is generally true \cite{Ibanez:1992hc}. It then follows from these equations that
\bea
&&{b_3'}^3-{b_2'}^3\in \mathbb{Z} \label{b32x}\\
&&\frac{5}{3}({b_3'}^3-{b_1'}^3) \in \mathbb{Z} \label{b31x}\\
&&\frac{5}{3}({b_3'}^3-{b_1'}^3)-({b_3'}^3-{b_2'}^3) \in 2\mathbb{Z} \label{b321x}
\eea
Evidently the half-integral solutions for ${b_3'}^3-{b_2'}^3$ in eqn (\ref{b321234}) are {\em not} consistent with the cancellation of modular anomalies. Further, if we {\em now} impose the constraint in eqn (\ref{512}), it follows that
\beq
{b_3'}^3-{b_2'}^3\in 2\mathbb{Z}
\eeq
so that only the earlier solutions in which ${b_3'}^3-{b_2'}^3=2$ can be consistent with modular anomaly cancellation.
 Eliminating the $\sum_{k}Y^k$ terms  from the above equations gives
\beq
6+\frac{5}{6}({b'}^3_{3}-{b_1'}^3) -\frac{1}{2}({b'}^3_{3}-{b_2'}^3)=\sum_{g,k=1}^3(2n^k_{Q_g}-n^k_{u_g}-n^k_{\ell_g}) \label{nQuellIL}
\eeq
and the modular weights displayed in Table \ref{n123} allow only the solutions displayed in Table \ref{Quell} when ${b_3'}^3-{b_2'}^3=2$ (and ${b_3'}^3-{b_1'}^3=24/5$). 
%%%%%%%%%%%%%%%%%%%%%%%%%%%%%%%%%
\begin{table}
 \begin{center}
\begin{tabular}{||c|c|c|c||} \hline \hline
$-\sum_{k,g=1}^3n^k_{Q_g}$&$-\sum_{k,g=1}^3n^k_{u_g}$&$-\sum_{k,g=1}^3n^k_{\ell_g}$& $-\sum_{k=1}^3 Y^k$\\ \hline \hline
3&6&9&11 \\
3&7&8&12 \\
3&8&7&13 \\
3&9&6&14 \\
4&8&9&12 \\
4&9&8&13 \\
\hline \hline
\end{tabular}
\end{center} 
\caption{ \label{Quell} Modular weights satisfying eqn (\ref{nQuellIL}) when ${b_3'}^3-{b_2'}^3=2$.}
 \end{table}
%%%%%%%%%%%%%%%%%%%%%%%%%%%%%%
However, we also need to satisfy eqns (\ref{b320}) and (\ref{b310}) with ${b_3'}^3-{b_2'}^3=2$. The only solutions are displayed in Table \ref{Quell5}. 
%%%%%%%%%%%%%%%%%%%%%%%%%%%%%%%%
\begin{table}
 \begin{center}
\begin{tabular}{||c|c|c|c||} \hline \hline
$-\sum_{g=1}^3n^3_{Q_g}$&$-\sum_{g=1}^3n^3_{u_g}$&$-\sum_{g=1}^3n^3_{\ell_g}$& $- Y^3$\\ \hline \hline
0&2&3&6 \\
0&3&2&7 \\
 0&$\frac{9}{4}$&$\frac{11}{4}$&$\frac{25}{4}$ \\
 0&$\frac{11}{4}$&$\frac{9}{4}$&$\frac{27}{4}$ \\
0&$\frac{5}{2}$&$\frac{5}{2}$&$\frac{13}{2}$\\
$\frac{1}{2}$&3&3&$\frac{13}{2}$\\
\hline \hline
\end{tabular}
\end{center} 
\caption{ \label{Quell5} Modular weights satisfying  eqns (\ref{b320}) and (\ref{b310}) when ${b_3'}^3-{b_2'}^3=2$.}
 \end{table}
%%%%%%%%%%%%%%%%%%%%%%%%%%%%%%
The question is whether any of these is achieved by a set of modular weights that also yields one of the solutions given in Table \ref{Quell}. 
Three of these require $\sum_{g=1}^3n^3_{u_g}=-3$ or $\sum_{g=1}^3n^3_{\ell_g}=-3$.  However this can only be achieved  for $u_g$ if $(n^1_{u_g},n^2_{u_g}n^3_{u_g})=(0,0,1)$ for all three generations. But then $\sum_{k,g=1}^3n^k_{u_g}=-3$ which is not allowed by any of the solutions given in Table \ref{Quell}. The same argument applies to $\ell_g$. 
 Similarly, there are two solutions in which $\sum_{g=1}^3n^3_{u_g}=-11/4$ or $\sum_{g=1}^3n^3_{\ell_g}=-11/4$, and the allowed sets of modular weights satisfying this have $\sum_{k,g=1}^3n^k_{u_g}=-4,-5$, neither of which is allowed  by the solutions in Table \ref{Quell}. It follows that the only possibility is  the solution with $\sum_{g=1}^3n^3_{u_g}=-5/2=\sum_{g=1}^3n^3_{\ell_g}$, which can arise from modular weights with $\sum_{k,g=1}^3n^k_{u_g}=-6,-7=\sum_{k,g=1}^3n^k_{\ell_g}$. Unfortunately none of the solutions in Table \ref{Quell} have {\em both} $\sum_{k,g=1}^3n^k_{u_g}=-6,-7$ and $\sum_{k,g=1}^3n^k_{\ell_g}=-6,-7$. The conclusion is that, independently of the moduli-dependence of the threshold corrections, it is impossible to satisfy the unification constraint given in eqn (\ref{512}) consistently with the cancellation of modular anomalies.

In their treatment of this question, instead of imposing eqn (\ref{512}),  Ib\'a\~nez and L\"ust \cite{Ibanez:1992hc} allowed a range of values of 
\beq
r:=\frac{\frac{5}{3}({b_3'}^3-{b_1'}^3)}{{b_3'}^3-{b_2'}^3}
\eeq
They required that the ratio
\beq
\gamma := \frac{1+r}{\frac{3}{5}r-1}
\eeq
is in the range
\beq
2.7 \leq \gamma \leq 3.7 
\eeq
corresponding to 
\beq
3.88 \leq r \leq 5.95 \label{rweak}
\eeq
which includes the value $r=4$ corresponding to eqn (\ref{512}). Eqns (\ref{b32x}) ... (\ref{b321x}) hold quite generally, and if we choose
\beq
{b_3'}^3-{b_2'}^3=1
\eeq
then the above range only allows $r=5$ consistently with eqn  (\ref{b321x}), in which case
\beq
{b_3'}^3-{b_1'}^3=3
\eeq
In this case the solutions of  eqns (\ref{b32y}) and (\ref{b31y}) are given in Table \ref{QuellIL2}, and the solutions of eqns (\ref{b320}) and (\ref{b310})  in Table \ref{Quell7}. 
%%%%%%%%%%%%%%%%%%%%%%%%%%%%%%%%%
\begin{table}
 \begin{center}
\begin{tabular}{||c|c|c|c||} \hline \hline
$-\sum_{i,g=1}^3n^i_{Q_g}$&$-\sum_{i,g=1}^3n^i_{u_g}$&$-\sum_{i,g=1}^3n^i_{\ell_g}$& $-\sum_{i=1}^3 Y^i$\\ \hline \hline
3&5&9&9 \\
3&6&8&10 \\
3&7&7&11 \\
3&8&6&12 \\
3&9&5&13 \\
4&7&9&10 \\
4&8&8&11 \\
4&9&7&12 \\
5&9&9&11 \\
\hline \hline
\end{tabular}
\end{center} 
\caption{ \label{QuellIL2} Modular weights satisfying eqns (\ref{b32y}) and (\ref{b31y})   when ${b'}^3_{3}-{b'}^3_{2}=1$.}
 \end{table}
%%%%%%%%%%%%%%%%%%%%%%%%%%%%%%
%%%%%%%%%%%%%%%%%%%%%%%%%%%%%%%%%%%%%%%%%%%%%%%%%%%%%%%%
\begin{table}
 \begin{center}
\begin{tabular}{||c|c|c|c||} \hline \hline
$-\sum_{g=1}^3n^3_{Q_g}$&$-\sum_{g=1}^3n^3_{u_g}$&$-\sum_{g=1}^3n^3_{\ell_g}$& $- Y^3$\\ \hline \hline
0&1&3&4 \\
0&$\frac{5}{4}$& $\frac{11}{4}$ & $\frac{17}{4}$ \\
0&$\frac{3}{2}$& $\frac{5}{2}$ & $\frac{9}{2}$ \\
0&$\frac{7}{4}$& $\frac{9}{4}$ & $\frac{19}{4}$ \\
0&2&2&5 \\
0&$\frac{9}{4}$& $\frac{7}{4}$ & $\frac{21}{4}$ \\
0&$\frac{5}{2}$& $\frac{3}{2}$ & $\frac{11}{2}$ \\
0&$\frac{11}{4}$& $\frac{5}{4}$ & $\frac{23}{4}$ \\
0&3&1&6\\
%%%%%%%%%%%
$\frac{1}{2}$&2&3 & $\frac{9}{2}$ \\
$\frac{1}{2}$&$\frac{9}{4}$& $\frac{11}{4}$ & $\frac{19}{4}$ \\
$\frac{1}{2}$&$\frac{5}{2}$& $\frac{5}{2}$ & 5\\
$\frac{1}{2}$&$\frac{11}{4}$& $\frac{9}{4}$ & $\frac{21}{4}$ \\
$\frac{1}{2}$&3&2 & $\frac{11}{2}$ \\
%%%%%%%%%%%%%%%%
$\frac{3}{4}$&$\frac{5}{2}$& 3& $\frac{19}{4}$ \\
$\frac{3}{4}$&$\frac{11}{4}$& $\frac{11}{4}$& 5 \\
$\frac{3}{4}$&$3$& $\frac{5}{2}$&  $\frac{21}{4}$\\
%%%%%%%%%%%%%%%%%%%%%%
1&3&3&5 \\
\hline \hline
\end{tabular}
\end{center} 
\caption{ \label{Quell7} Modular weights satisfying eqns   (\ref{b320}) and (\ref{b310})  when $b_{3}'-b_{2}'=1$.}
 \end{table}
%%%%%%%%%%%%%%%%%%%%%%%%%%%%%%%%%%%%%%%%%%%%%%%%%%%%%%%%%%%%%%%%%%%%%%%%%%%%%%%%%%%%%%%%%%%%%%%%%%%%%%%%%%%%%%
The solution given in  \cite{Ibanez:1992hc} corresponds to the third line in each Table, and it is easily verified that their solution satisfies the modular anomaly cancellation constraints given in eqn (\ref{modanom}). Thus the $\mathbb{Z}_8$-I orbifold does allow gauge coupling constant unification if the weaker constraint given in eqn (\ref{rweak}) is imposed rather than $r=4$ that is required by eqn (\ref{512}). The outstanding question then is whether this can be achieved with a reasonable value of the K\"ahler modulus $T_3$ of order unity. We shall address this question in  \S \ref{KT}.
%%%%%%%%%%%%%%%%%%%%%%%%%%%%%%%%%%%%%%%%%%%%%%%%%%%%%%%%%%%%%%%%%%%%%%%%%%%%%
\subsection{The $\mathbb{Z}_3 \times \mathbb{Z}_6$ orbifold}
The point group generators $\theta, \omega$ have eigenvalues $e^{2\pi i v_k}$ with
\bea 
 (v_1,v_2,v_3)_{\theta}&&= \frac{1}{3}(0,-1,1) \\
(v_1,v_2,v_3)_{\omega}&&= \frac{1}{6}(1,-1,0) 
\eea
which may be realised using Casimirs on the $SU(3)^3$ lattice. It is convenient to reorder the twists here compared with those used in \cite{Love:1996sk} and Table \ref{ZN} of \S2, so that now $T^2_3$ is the $\mathbb{Z}_3$-plane, while $T^2_{1,2}$ are both $\mathbb{Z}_6$-planes. 
Evidently all three planes $T^2_{1,2,3}$ are $\mathcal{N}=2$ planes, so all three K\"ahler moduli $T_{1,2,3}$ contribute to the threshold corrections. The complex structure moduli $U_{1,2,3} =e^{2\pi i/ 3}$ are all fixed, since none of them is a $\mathbb{Z}_2$-plane. 
In this case eqns (\ref{Del32})  and  (\ref{Del321}) require that 
\bea
\sum_{k=1}^3({b'_{3}}^k-{b'_{2}}^k) \ln \alpha _k <0  \label{3632}\\
\sum_{k=1}^3({b'_{3}}^k-{b'_{1}}^k) \ln \alpha _k <0 \label{3631}\\
\frac{\sum_{k=1}^3({b'_{3}}^k-{b'_{2}}^k) \ln \alpha _k }{\sum_{k=1}^3({b'_{3}}^k-{b'_{1}}^k) \ln \alpha _k}=\frac{5}{12} \label{512alph}
\eea
where 
\beq
\alpha _k :=(T_k+\bar{T}_k)|\eta(p_kT_k)\eta(q_kT_k)|^2
\eeq
The quantities $\ln \alpha_k$ are dominated the contribution from the Dedekind eta functions whose logarithm is negative in the region of interest, as can be seen in eqn (\ref{etaT}); this explains the difference between the 
 inequalities in eqns (\ref{3632}) and (\ref{3631}) compared with those  in eqns (\ref{b32}) and (\ref{b31}).
If a single modulus, $T_3$ say, dominates, these equations reduce to the constraints given previously in eqns (\ref{b320}), (\ref{b310}) and (\ref{512}). Further, there are no modular anomalies on the $\mathbb{Z}_3 \times \mathbb{Z}_6$ orbifold. In this case the modular weights \cite{Ibanez:1992hc,Bailin:1992hx} are given in Table \ref{n12336}, and using these it is clear from eqn (\ref{b320}) that 
$3({b'_{3}}^3-{b'_{2}}^3)\in \mathbb{Z}$. In fact, using eqn (\ref{512}), it follows from eqn (\ref{sumgn35a}) that $3({b'_{3}}^3-{b'_{2}}^3)\in 2\mathbb{Z}$. Table \ref{QuellY36} lists all solutions with ${b'_{3}}^3-{b'_{2}}^3>0$, and shows that the only allowed values are  $3({b'_{3}}^3-{b'_{2}}^3)=2,4,6,8$; it is easy to see that there are many allowed choices of the modular weights $n^3_{({h,\bar{h}, d,L})_g} $ that yield the values of $Y^3$ that arise.
%%%%%%%%%%%%%%%%%%%%%%%%%%%%%%%%%%%%%%%%%%%%%%%%%%%%%%%%%%%%%%%%%%%%%%%
\begin{table}
 \begin{center}
\begin{tabular}{||c||c|c|c|c||} \hline \hline
State $X$ & $-n_X^1$ & $-n_X^2$ & $-n_X^3$ &$-\sum_i n^i_X$ \\ \hline \hline
 $Q, u, \ell$ \ {\rm and} \ $L,h,\bar{h}, d$&$\frac{5}{6}$& -$\frac{5}{6}$&0&0 \\
&-$\frac{5}{6}$& $\frac{5}{6}$&0 &0\\ \hline
&1&0&0 &1\\
&0&1&0&1 \\
&0&0&1&1 \\
&$\frac{5}{6}$& $\frac{1}{6}$&0 &1\\
&$\frac{1}{6}$& $\frac{5}{6}$&0 &1\\
&$\frac{1}{3}$& $\frac{2}{3}$&0 &1\\
&$\frac{2}{3}$& $\frac{1}{3}$&0 &1\\
&$\frac{1}{2}$& $\frac{1}{2}$&0 &1\\
&0&$\frac{1}{3}$& $\frac{2}{3}$ &1\\
&0&$\frac{2}{3}$& $\frac{1}{3}$ &1\\
&$\frac{1}{3}$&0& $\frac{2}{3}$ &1\\
&$\frac{2}{3}$&0& $\frac{1}{3}$ &1\\ \hline
&$\frac{1}{6}$& $\frac{11}{6}$&0 &2\\
&$\frac{11}{6}$& $\frac{1}{6}$&0 &2\\
&$\frac{5}{6}$& $\frac{5}{6}$&$\frac{1}{3}$ &2\\ 
&$\frac{5}{6}$& $\frac{1}{2}$&$\frac{2}{3}$ &2\\ 
&$\frac{2}{3}$& $\frac{2}{3}$&$\frac{2}{3}$ &2\\ 
&$\frac{1}{2}$& $\frac{5}{6}$&$\frac{1}{3}$ &2\\ 
\hline \hline
%%%%%%
{\rm Only} \ $L,h,\bar{h}, d$ &-$\frac{11}{6}$ & $\frac{5}{6}$ &0&-1 \\
&-$\frac{11}{6}$ & $\frac{5}{6}$ &0&-1 \\ \hline
 &-$\frac{1}{3}$& $\frac{1}{3}$&0 &0\\
 &$\frac{1}{3}$& $\frac{1}{3}$&0 &0\\
&0&-$\frac{1}{3}$& $\frac{1}{3}$ &0\\
 &0&$\frac{1}{3}$& -$\frac{1}{3}$ &0\\
&-$\frac{1}{3}$&0& $\frac{1}{3}$ &0\\
 &$\frac{1}{3}$&0& -$\frac{1}{3}$ &0\\ \hline
 &$\frac{1}{3}$ & $\frac{5}{3}$ &0&2 \\
&$\frac{5}{3}$ & $\frac{1}{3}$ &0&2 \\
&0&$\frac{1}{3}$ & $\frac{5}{3}$ &2 \\
&0&$\frac{5}{3}$ & $\frac{1}{3}$ &2 \\
&$\frac{1}{3}$ &0& $\frac{5}{3}$ &2 \\
&$\frac{5}{3}$ &0& $\frac{1}{3}$ &2 \\ \hline
&$\frac{11}{6}$ & $\frac{5}{6}$ &$\frac{1}{3}$&3 \\
&$\frac{5}{6}$ & $\frac{11}{6}$ &$\frac{1}{3}$&3 \\
&$\frac{11}{6}$ & $\frac{1}{2}$ &$\frac{2}{3}$&3 \\
&$\frac{1}{2}$ & $\frac{11}{6}$ &$\frac{2}{3}$&3 \\ \hline
%%%%%%%%%%%%%%
&$\frac{17}{6}$ & $\frac{5}{6}$ &$\frac{1}{3}$&4 \\
&$\frac{5}{6}$ & $\frac{17}{6}$ &$\frac{1}{3}$&4 \\
&$\frac{11}{6}$ & $\frac{11}{6}$ &$\frac{1}{3}$&4 \\
&$\frac{5}{3}$ & $\frac{5}{3}$ &$\frac{2}{3}$&4 \\
\hline \hline
\end{tabular}
\end{center} 
\caption{ \label{n12336} Modular weights of matter representations on the $\mathbb{Z}_3 \times\mathbb{Z}_6$ orbifold. %The labels $Q_g$ and $L_g$ respectively are the quark  and lepton doublets,  $u_g,d_g$ and $\ell_g$ the quark and lepton singlets, and $h,\bar{h}$ are the two Higgs doublets;
}
 \end{table}
%%%%%%%%%%%%%%%%%%%%%%%%%%%%%%%%%%%%%%%%%%%%%%%%%%%%%%%%%%%%%%%%%%%%%%%%%%%%%%%%%%%%%%%%%%%%%%%%%%
%%%%%%%%%%%%%%%%%%%%%%%%%%%%%%%%%%%%%%%%%%%%%%%%%%%%%%%%%%%%%%%%%%%%%%%
\begin{table}
 \begin{center}
\begin{tabular}{||c||c|c|c|c|c||} \hline \hline
${b'_{32}}^3$&$-\sum_{g}3n^3_{Q_g}$&$-\sum_{g}3n^3_{u_g}$&$-\sum_{g}3n^3_{\ell_g}$&$-3Y^3$&Table \ref{QuellY} solution \\ \hline \hline
%%%%%%%%%%%%%%%%%%%%%%%%%%%%%
$\frac{2}{3}$&0&$(0,1,...,9)$&$(9,8,...,0)$&$(8,9,...,17)$&$=$II$\supset$ I$\supset$ III $\supset$ IV\\
&1&$(2,3,...,9)$&$(9,8,...,2)$&$(9,10,...,16)$& $\supset$ III\\
&2&$(4,5,...,9)$&$(9,8,...,4)$&$(10,11,...,15)$&$\supset$ III\\
&3&$(6,7,8,9)$&$(9,8,7,6)$&$(11,12,13,14)$&\\
&4&$(8,9)$&$(9,8)$&$(12,13)$&\\ \hline
%%%%%%%%%%%%%%%%%%%%%%%%%%%%
$\frac{4}{3}$&0&$(3,4,...,9)$&$(9,8,...,3)$&$(13,14,...,19)$&$=$II$\supset$ I$\supset$ III $\supset$ IV\\
&1&$(5,6,...,9)$&$(9,8,...,5)$&$(14,15,...,18)$&\\
&2&$(7,8,9)$&$(9,8,7)$&$(15,16,17)$&\\
&3&$9$&$9$&$16$&\\ \hline
%%%%%%%%%%%%%%%%%%%%%%%%%%%%%
2&0&$(6,7,8,9)$&$(9,8,7,6)$&$(18,19,20,21)$&$=$ II $\supset$ I\\
&1&$(8,9)$&$(9,8)$&$(19,20)$&\\ \hline
%%%%%%%%%%%%%%%%%%%%%%%%%%%%%%%%%%%%%%%
$\frac{8}{3}$&0&9&9&23&\\
\hline \hline
\end{tabular}
\end{center} 
\caption{ \label{QuellY36} All solutions of eqn (\ref{sumgn35a}) with ${b'_{3}}^3-{b'_{2}}^3>0$.The last column indicates which solutions in Table  \ref{QuellY}  wholly or partially realise this solution here.}
 \end{table}
%%%%%%%%

However, it is probably unrealistic to assume that a single modulus dominates on this orbifold. Instead we assume that 
\beq
T_1=T_2=T_3:=T
\eeq
We wish to exploit the considerably reduced modular symmetry that may be associated with the K\"ahler modulus $T_3$ when there are Wilson lines in this $\mathbb{Z}_3$ plane. Thus we choose $\alpha_3 \neq \alpha_{1}=\alpha_2$; $T^2_{1,2}$ are both $\mathbb{Z}_6$ planes, so it is perfectly feasible that their modular symmetries are the same (even if there are non-trivial Wilson lines) and that $\alpha_{1}=\alpha_2$. 
Then eqn (\ref{512alph}) gives
\beq
\frac{\sum_{k=1}^2({b'_{3}}^k-{b'_{2}}^k) \ln \alpha _1+({b'_{3}}^3-{b'_{2}}^3)\ln \alpha _3}{\sum_{k=1}^2({b'_{3}}^k-{b'_{1}}^k) \ln \alpha _1+({b'_{3}}^3-{b'_{2}}^3)\ln \alpha _3}=\frac{5}{12}
\eeq
where
\beq
\alpha_1=\alpha_2=(T+\bar{T})|\eta(p_1T)\eta(q_1T)|^2
\eeq
This is satisfied independently of $\alpha _1$ and $\alpha _3$  if, 
\beq
\frac{\sum_{k=1}^2({b'_{3}}^k-{b'_{2}}^k)}{\sum_{k=1}^2{({b'_{3}}^k-{b'_{1}}^k)}}=\frac{5}{12}=
\frac{      {b'_{3}}^3   -   {b'_{2}}^3    }{    {b'_{3}}^3  -   {b'_{1}}^3    } \label{512p}
\eeq
and this implies that
\beq
\frac{\sum_{k=1}^3({b'_{3}}^k-{b'_{2}}^k)}{\sum_{k=1}^3{({b'_{3}}^k-{b'_{1}}^k)}}=\frac{5}{12} \label{512q}
\eeq
We have just given (in Table \ref{QuellY36}) all solutions of the second equation in  equation (\ref{512p}). So the question is whether there are solutions of eqn (\ref{512q}) that are achievable with the same sets of modular weights as are needed to solve the former.

Summing over $k=1,2,3$, it follows, as in eqns (\ref{b32y}) and (\ref{b31y}), that
\bea
 &&\sum_{k=1}^3({b'_{3}}^k-{b'_{2}}^k)=-6-\sum_{k=1}^3Y^k+\sum_{g,k=1}^3(-n^k_{Q_g}+n^k_{u_g}) \\
&&\frac{5}{3}\sum_{k=1}^3({b'_{3}}^k-{b'_{1}}^k)=-18-\sum_{k=1}^3Y^k+\sum_{g,k=1}^3(3n^k_{Q_g}-n^k_{u_g}-2n^k_{\ell_g})
\eea
where $Y_k$ is given in eqn (\ref{Xi}). 
 As before, we may infer, quite generally, that
\bea
&&\sum_{k=1}^3({b'_{3}}^k-{b'_{2}}^k) \in \mathbb{Z} \\
&&\frac{5}{3}\sum_{k=1}^3({b'_{3}}^k-{b'_{1}}^k) \in \mathbb{Z} \\
&&\frac{5}{3}\sum_{k=1}^3({b'_{3}}^k-{b'_{1}}^k)-\sum_{k=1}^3({b'_{3}}^k-{b'_{2}}^k) =0 \bmod 2
\eea
Imposing eqn (\ref{512q}) then gives
\beq
3\sum_{k=1}^3({b'_{3}}^k-{b'_{2}}^k)=-12+\sum_{g,k=1}^3(4n^k_{Q_g}-2n^k_{u_g}-2n^k_{\ell})  \label{3b32}
\eeq
Thus, in this case 
\bea
&&\sum_{k=1}^3({b'_{3}}^k-{b'_{2}}^k)=0 \bmod 2 \\
&&\sum_{g,k=1}^3(n^k_{Q_g}+n^k_{u_g}+n^k_{\ell})=0\bmod 3
\eea
and using the modular weights in Table \ref{n12336} the only solutions are given in Table \ref{QuellY}. 
%%%%%%%%%%%%%%%%%%%%%%%%%%%%%%%%%%%%%%%%%%%%%%%%%%%%%%%%%%%%%%%%%%%%%%%
\begin{table}
 \begin{center}
\begin{tabular}{||c|c||c|c|c|c||} \hline \hline
Solution &$\sum_{k=1}^3({b'_{3}}^k-{b'_{2}}^k)$&$-\sum_{g,k=1}^3n^k_{Q_g}$&$-\sum_{g,k=1}^3n^k_{u_g}$&$-\sum_{g,k=1}^3n^k_{\ell_g}$&$\sum_{k=1}^3Y_k$ \\ \hline \hline
Ia&2&0&5&4&13 \\
Ib&2&0&4&5&12 \\
IIa&2&0&6&3&14 \\
IIb&2&0&3&6&11 \\
IIIa&2&1&6&5&13 \\
IIIb&2&1&5&6&12\\
IV&4&0&6&6&16 \\ 
\hline \hline
\end{tabular}
\end{center} 
\caption{ \label{QuellY} All solutions of eqn (\ref{3b32}) with $\sum_{k=1}^3({b'_{3}}^k-{b'_{2}}^k)>0$.}
 \end{table}
%%%%%%%%%%%%%%%%%%%%%%%%%%%%%%%%%%%%%%%%%%%%%%%%%%%%%%%%%%%%%%%%%%%%%%%%%%%%%%%%%%%%%%%%%%%%%%%%%%
As before, there are many sets of modular weights that give rise to the  values found of $\sum_{k=1}^3Y_k$. 

We first need to know whether any of the solutions displayed in Tables \ref{QuellY36} and \ref{QuellY} coincide.
Consider solution IV of Table \ref{QuellY}. From Table \ref{n12336}, we see that $\sum_{g,k=1}^3n^k_{Q_g}=0$ requires that $n^3_{Q_g}=0$ for all $g$, so this line can only be consistent with some or all of the  four lines in Table \ref{QuellY36} in which $\sum_{g=1}^3n^3_{Q_g}=0$. It also follows from Table \ref{n12336} that $-\sum_{g,k=1}^3n^k_{u_g}=6$ can only be realised  with $-\sum_{k=1}^3n^k_{u_g}=2$ for all $g$, and this requires that $-3n^3_{u_g}=(0,1,2)$ so that $-\sum_{g=1}^33n^3_{u_g}=(0,1,2, ..., 6)$. Similarly for $-\sum_{g=1}^33n^3_{\ell_g}$. Consider the solutions in the first line of Table \ref{QuellY36} with ${b'_{3}}^3-{b'_{2}}^3=2/3$. They have the property that $-\sum_{g}3n^3_{u_g}-\sum_{g}3n^3_{\ell_g}=9$. Thus only the four combinations $(-\sum_{g}3n^3_{u_g},-\sum_{g}3n^3_{\ell_g})=((3,4,5,6),(6,5,4,3))$ can be realised in this case. The values of $3Y_3$ corresponding to these four solutions are $3Y_3=(10,11,12,13)$.
Likewise, only one of the solutions in the line with ${b'_{3}}^3-{b'_{2}}^3=4/3$ may be realised, namely the one with  $(-\sum_{g}3n^3_{u_g},-\sum_{g}3n^3_{\ell_g})=(6,6)$. The solutions with  ${b'_{3}}^3-{b'_{2}}^3=2$ and $8/3$ cannot be realised from solution IV.
In summary, using solution IV we can realise only the solutions in Table \ref{QuellY36} in which 
\bea
\left({b'_{3}}^3-{b'_{2}}^3,-\sum_{g}3n^3_{Q_g},-\sum_{g}3n^3_{u_g},-\sum_{g}3n^3_{\ell_g},3Y_3\right)&&=\left(\frac{2}{3}, 0,(3,4,5,6),(6,5,4,3),(10,11,12,13)\right) \nonumber\\
&& \label{353}\\
&&=\left(\frac{4}{3},0,6,6,16\right)
\eea
%Eqn (\ref{353}) defines four solutions, all having the same values of $({b_3'}^3-{b_3'}^2, -\sum_{g}3n^3_{Q_g})=(2/3,0)$ but different values of $(-\sum_{g}3n^3_{u_g},-\sum_{g}3n^3_{\ell_g},3Y_3)$; in the first solution, for example, these quantities take the values $(3,6,10)$.
Also, as above,  solutions Ia,Ib,IIa,IIb of Table \ref{QuellY} can only be consistent with some or all of the  four lines in Table \ref{QuellY36} in which $\sum_{g=1}^3n^3_{Q_g}=0$. Proceeding as above, for solution Ia, we may infer that for two generations, $g=1,2$ say,  $-\sum_{k=1}^3n^k_{u_g}=2$, whereas for the third $-\sum_{k=1}^3n^k_{u_3}=1$. In this last case, $-3n^3_{u_3}=(0,1,2, 3)$, so that for  solution Ia $-\sum_{g=1}^33n^3_{u_g}=(0,1,2, ..., 7)$. It follows similarly that $-\sum_{g=1}^33n^3_{\ell_g}=(0,1,2, ..., 8)$; for solution Ib, the ranges for $u$ and $\ell$ are interchanged. Thus  now we can realise the solutions in Table \ref{QuellY36} in which
\bea
\left({b'_{3}}^3-{b'_{2}}^3,-\sum_{g}3n^3_{Q_g},-\sum_{g}3n^3_{u_g},-\sum_{g}3n^3_{\ell_g},3Y_3\right)&&=\left(\frac{2}{3}, 0,(1,2,...,8),(8,7,...,1),(9,10,...,16)\right) \nonumber\\
&& \label{230}\\
&&=\left(\frac{4}{3},0,(5,6,7),(7,6,5),16\right)\\
&&=\left(2,0,(7,8),(8,7),(19,20)\right)
\eea
Proceeding in this way, we find that the solutions in Table \ref{QuellY}  generate some of the solutions in  Table \ref{QuellY36}, but not all. Which ones are are displayed in Table \ref{consistent}, and for comparison in the last column of Table \ref{QuellY36}.
%%%%%%%%%%%%%%%%%%%%%%%%%%%%%%%%%%%%%%%%%%%%%%%%%%%%%%%%%%%%%%%%%%%%%%%
%%%%%%%%%%%%%%%%%%%%%%%%%%%%%%%%%%%%%%%%%%%%%%%%%%%%%%%%%%%%%%%%%%%%%%%
\begin{table}
 \begin{center}
\begin{tabular}{||c|c|c|c|c|c||} \hline \hline
Solution&${b'_{3}}^k-{b'_{2}}^3$&$-\sum_{g}3n^3_{Q_g}$&$-\sum_{g}3n^3_{u_g}$&$-\sum_{g}3n^3_{\ell_g}$&$-3Y^3$ \\ \hline \hline
%%%%%%%%%%%%%%%%%%%%%%%%%%%
I&$\frac{2}{3}$&0&$(1,2,...,8)$&$(8,7,...,1)$&$(9,10,...,16)$ \\
&$\frac{4}{3}$&0&$(4,5,...,8)$&$(8,7,...,4)$&$(14,15,...,18)$ \\
&$2$&0&$(7,8)$&$(8,7)$&$(19,20)$ \\ \hline
%%%%%%%%%%%%%%%%%%%%%%%%%
II&$\frac{2}{3}$&0&$(0,1,...,9)$&$(9,8,...,0)$&$(8,9,...,17)$ \\
&$\frac{4}{3}$&0&$(3,4,...,9)$&$(9,8,...,3)$&$(13,14,...,19)$ \\
&$2$&0&$(6,7,8,9)$&$(9,8,7,6)$&$(18,19,20,21)$ \\ \hline
%%%%%%%%%%%%%%%%%%%%%%%%%
III&$\frac{2}{3}$&0&$(2,3,...,7)$&$(7,6,...,2)$&$(10,11,...,15)$ \\
&$\frac{4}{3}$&0&$(5,6,7)$&$(7,6,5)$&$(15,16,17)$ \\
&$\frac{2}{3}$&1&$(4,5,6,7)$&$(7,6,5,4)$&$(11,12,13,14)$ \\
&$\frac{2}{3}$&2&$(6,7)$&$(7,6)$&$(12,13)$ \\ \hline
%%%%%%%%%%%%%%%%%%%%%%%%%
IV&$\frac{2}{3}$&0&$(3,4,5,6)$&$(6,5,4,3)$&$(12,13,14,15)$ \\
&$\frac{4}{3}$&0&6&6&16 \\ \hline \hline
\end{tabular}
\end{center} 
\caption{ \label{consistent} Consistent solutions from Table \ref{QuellY}.}
 \end{table}
%%%%%%%%

Finally, we need to check that the consistent solutions may all be realised by consistent sets of modular weights for $h, \bar{h}, d_g, L_g$. For example, consider solution Ia in Table \ref{QuellY}. The particular choice of  the modular weights displayed in Table \ref{Ia} realises the values $(2/3,0,4,5,12)$ on the right-hand side of eqn (\ref{230}) and satisfies all of our constraints. The other solutions may be similarly realised.
%%%%%%%%%%%%%%%%%%%%%%%%%%%%%%%%%%%%%%%%%%%%%%%%%%%%%%%%%%%%%%%%%%%%%%%
\begin{table}
 \begin{center}
\begin{tabular}{||c||c|c|c|c||} \hline \hline
Particle $X$ & $-n^1_X$ & $-n^2_X$ & $-n^3_X$ & $-\sum_kn^k_X$ \\ \hline \hline
$h$& $\frac{11}{6}$ &$\frac{5}{6}$& $\frac{1}{3}$& $3$ \\
$\bar{h}$& $\frac{11}{6}$ &$\frac{1}{2}$& $\frac{2}{3}$& $3$ \\
$L_1$&$0$ & $\frac{1}{3}$&$\frac{5}{3}$&2 \\
$L_2$&$\frac{2}{3}$ & $\frac{2}{3}$&$\frac{2}{3}$&2 \\
$L_3$&$\frac{11}{6}$ &$\frac{1}{2}$& $\frac{2}{3}$& $3$ \\
$d_1$& $\frac{5}{6}$& $-\frac{5}{6}$&0&0 \\ 
$d_2$& $\frac{5}{6}$& $-\frac{5}{6}$&0&0 \\ 
$d_3$& $\frac{5}{6}$& $-\frac{5}{6}$&0&0 \\ \hline
& $-Y^1=\frac{11}{3}$& $-Y^2=\frac{16}{3}$ &$-Y^3= 4$&$-\sum_kY^k=13$ \\ \hline
$Q_1$&$\frac{5}{6}$ &$-\frac{5}{6}$&0&0 \\
$Q_2$&$\frac{5}{6}$ &$-\frac{5}{6}$&0&0 \\
$Q_3$&$\frac{5}{6}$ &$-\frac{5}{6}$&0&0 \\
$u_1$&$\frac{5}{6}$ &$\frac{5}{6}$&$\frac{1}{3}$&2 \\
$u_2$&$\frac{5}{6}$ &$\frac{5}{6}$&$\frac{1}{3}$&2 \\
$u_3$&$\frac{1}{3}$ &0&$\frac{2}{3}$&1 \\
$\ell_1$& 0 & $\frac{2}{3}$& $\frac{1}{3}$&1 \\
$\ell_2$& 0 & $\frac{1}{3}$& $\frac{2}{3}$&1 \\
$\ell_3$& $\frac{2}{3}$ & $\frac{2}{3}$& $\frac{2}{3}$&1 \\ \hline
& ${b'_{32}}^1=\frac{13}{6}$& ${b'_{32}}^2=-\frac{5}{6}$& ${b'_{32}}^3=\frac{2}{3}$& $\sum_k{b'_{32}}^k=2$ \\

& ${b'_{31}}^1=-\frac{39}{10}$& ${b'_{31}}^2=\frac{71}{10}$& ${b'_{31}}^3=\frac{8}{5}$& $\sum_k{b'_{31}}^k=\frac{24}{5}$ \\
\hline \hline
\end{tabular}
\end{center} 
\caption{ \label{Ia} Consistent modular weights for solution Ia.}
 \end{table}
%%%%%%%%
%%%%%%%%%%%%%%%%%%%%%%%%%%%%%%%%%%%%%%%%%%%%%%%%%%%%%%%%%%%%%%%%%%%%%%%%%%%%%%%%%%%%%%%%%%%%%%%%%%%%%%%%%%%%%%%%%%%%%%%%%%%%%%%%%%%%%%%%%%%%%%%%%%%%%%
%%%%%%%%%%%%%%%%%%%%%%%%%%%%%%%%%%%%%%%%%%%%%%%%%%%%%%
\section{Estimation of the K\"ahler modulus needed for unification} \label{KT}
It follows from eqn (\ref{ms2mx2}) that
\bea
-2(b_3-b_2)\ln\frac{m_{\rm string}}{m_X}&&=\Delta_3-\Delta_2 \\
&&=-\sum_{k=1}^3({b'_3}^k-{b'_2}^k)\ln [(T_k+\bar{T}_k)|\eta(p_kT_k)\eta(q_kT_k)|^2]
\eea
For all values of interest we may approximate
\beq
\ln(\eta(T)) \simeq -\frac{\pi T}{12}  \label{etaT}
\eeq
Thus, since $b_3-b_2=-4$ and $\ln(m_{\rm string}/m_X) \simeq 2.92$, in general, we seek allowed values of $p_k+q_k$ for which there are solutions of 
\beq
23.3\simeq \sum_{k=1}^3({b'_3}^k-{b'_2}^k)\left[\frac{\pi (p_k+q_k) T_k}{6}-\ln (2T_k)\right] \label{uni3}
\eeq
with $T_k \sim 1$.
%%%%%%%%%%%%%%%%%%%%%%%%%%%%%%%%%%%%%%%%%%%%%%%%%%
\subsection{Unification in the $\mathbb{Z}_8$-I orbifold}
In the case of the $\mathbb{Z}_8$-I orbifold the threshold corrections are determined entirely by the modulus $T_3$. As discussed in \S 3.1, it is impossible to satisfy the ``strong'' unification constraint given in eqn (\ref{512}). However if instead, as in \cite{Ibanez:1992hc}, we  impose the weaker constraint given in eqn (\ref{rweak}), then there is an allowed solution in which ${b_3'}^3-{b_2'}^3=1$, corresponding to
\beq
\frac{{b_3'}^3-{b_2'}^3}{{b_3'}^3-{b_1'}^3}=\frac{4}{12} \label{412}
\eeq
In this case, then, we must satisfy
\beq
23.3\simeq  \frac{\pi (p_3+q_3) T_3}{6}-\ln (2T_3) \label{uni}
\eeq
In \S 2.2 we showed that, depending on  the choice of Wilson lines,  the modular symmetry  associated with a $\mathbb{Z}_4$-plane may be $\Gamma_0(2), \Gamma_0(4)$ or $\Gamma_0(16)$. If we choose Wilson lines ${\bf A}_{5}$ and ${\bf A}_{6}$ such that ${\bf A}_{5}^tC{\bf A}_{6}\in \mathbb{Z}+\frac{1}{2}$, it is  {\em plausible} that the threshold corrections involve terms with
\beq
p_3=16=q_3
\eeq                    
in which case eqn (\ref{uni}) is satisfied when
\beq
T_3\simeq 1.5
\eeq
%%%%%%%%%%%%%%%%%%%%%%%%%%%%%%%%%%%%%%%%%%%%%%%%%%%%%%%%%%%%%%%
\subsection{Unification in the $\mathbb{Z}_3 \times \mathbb{Z}_6$ orbifold}
As discussed in \S 3.2, all three K\"ahler moduli contribute to the threshold corrections in this case. However, if the  modulus $T_3$ associated with the $\mathbb{Z}_3$-plane dominates,  then 
 it is certainly possible to satisfy the strong unification constraint. The allowed values of  ${b_3'}^3-{b_2'}^3$, given in Table \ref{QuellY36}, are
\beq
{b_3'}^3-{b_2'}^3=\frac{1}{3}(2,4,6,8)
\eeq
If we choose Wilson lines ${\bf A}_{5}$ and ${\bf A}_{6}$ in $T^2_3$ that are not lattice vectors such that $3{\bf A}_{5}^tC{\bf A}_{6}\in 2\mathbb{Z}$, then the modular group associated with the $\mathbb{Z}_3$-plane is $\Gamma_0(9)$, whereas if $3{\bf A}_{5}^tC{\bf A}_{6}\in 2\mathbb{Z}+1$, then it is $\Gamma_0(36)$. Table \ref{z36Tvalues} shows how it is possible to satisfy the strong unification constraint with $T_3 \sim 1$ for all allowed values of ${b_3'}^3-{b_2'}^3$ by a judicious, but plausible, choice of $p_3$ and $q_3$.
%%%%%%%%%%%%%%%%%%%%%%%%%%%%%%%%%%%%%%%%%%%%%%%%%%%%%%%%%%%%%%%%%%%%%%%
\begin{table}
 \begin{center}
\begin{tabular}{||c||c|c|c||} \hline \hline
 ${b_3'}^3-{b_2'}^3$ & $p_3$ & $q_3$ & $T_3$ \\ \hline \hline
$\frac{2}{3}$&36&36&0.94 \\
$\frac{4}{3}$&36&1&0.94 \\
2&9&9&1.34 \\
$\frac{8}{3}$&9&9&1.0 \\ \hline \hline
\end{tabular}
\end{center} 
\caption{ \label{z36Tvalues} Choices of $p_3$ and $q_3$ for the $\mathbb{Z}_3 \times \mathbb{Z}_6$ orbifold that allow unification with $T_3\sim 1$.}
 \end{table}
%%%%%%%%%%%%%%%%%%%%%%%%%%%%%%%%%%%%%%%%%%%%%%%%%%%%%%%%%%%%%%%%%%%%%%%%%%%

It is more natural to assume that all three moduli  contribute and are of the same size $T_1 \simeq T_2 \simeq T_3 =T$, and we have shown in Table \ref{consistent} that the strong unification condition can still be satisfied, but only for the combinations of ${b_3'}^3-{b_2'}^3$ and $\sum_{k=1}^3({b_3'}^3-{b_2'}^k)$ given in the first and third columns of Table \ref{z363Tvalues}. 
Wilson lines ${\bf A}_{2k-1}$ and ${\bf A}_{2k} $ in the $\mathbb{Z}_6$-planes $T^2_k \ (k=1,2)$ are always $E_8 \times E_8$ lattice vectors.  If ${\bf A}_{2k-1}^tC{\bf A}_{2k} \in 2\mathbb{Z}$, then the associated modular symmetry is the full $PSL(2 ,\mathbb{Z})$. Alternatively,  if ${\bf A}_{2k-1}^tC{\bf A}_{2k} \in 2\mathbb{Z}+1$, then it is $\Gamma_0(4)$. We choose the Wilson lines such that the symmetry is the same in both $\mathbb{Z}_6$-planes, and $\alpha_1=\alpha_2$. Then eqn (\ref{uni3}) becomes
\beq
23.3\simeq \frac{\pi T}{6}\left[ ({b'_3}^3-{b'_2}^3)(p_3+q_3)+\sum_{k=1}^2({b'_3}^k-{b'_2}^k)(p_1+q_1)\right]-\ln (2T)\sum_{k=1}^3({b'_3}^k-{b'_2}^k) \label{uni31}
\eeq
Table \ref{z363Tvalues} shows how it is possible to satisfy the strong unification constraint with $T \sim 1$ for all allowed values of $({b_3'}^3-{b_2'}^3,\sum_{k=1}^3({b'_3}^k-{b'_2}^k))$ by a judicious, but plausible, choice of $p_3,q_3, p_1$ and $q_1$. (The solution in the last line is the same as the solution in the third line of Table \ref{z36Tvalues} because the moduli associated with $T^2_1$ and $T^2_2$ do not contribute; in other words, the threshold corrections are dominated by $T_3$ in this case.)
%%%%%%%%%%%%%%%%%%%%%%%%%%%%%%%%%%%%%%%%%%%%%%%%%%%%%%%%%%%%%%%%%%%%%%%
\begin{table}
 \begin{center}
\begin{tabular}{||c|c|c||c|c|c|c|c||} \hline \hline
 ${b_3'}^3-{b_2'}^3$ &$\sum_{k=1}^2({b'_3}^k-{b'_2}^k)$&$\sum_{k=1}^3({b_3'}^3-{b_2'}^k)$& $p_3$ & $q_3$ &$p_1$&$q_1$& $T$ \\ \hline \hline
$\frac{2}{3}$&$\frac{4}{3}$ & 2&36&36&1&1&0.92 \\
$\frac{2}{3}$&$\frac{10}{3}$ & 4&36&36&1&1&0.9 \\
$\frac{4}{3}$&$\frac{2}{3}$ & 2&36&1&1&1&0.92 \\
$\frac{4}{3}$&$\frac{8}{3}$ & 4&36&1&1&1&0.9 \\
2&0&2&9&9&-&-&1.34\\
 \hline \hline
\end{tabular}
\end{center} 
\caption{ \label{z363Tvalues} Choices of $p_3,q_3, p_1$ and $q_1$ for the $\mathbb{Z}_3 \times \mathbb{Z}_6$ orbifold  that allow unification with $T\sim 1$.}
 \end{table}
%%%%%%%%%%%%%%%%%%%%%%%%%%%%%%%%%%%%%%%%%%%%%%%%%%%%%%%%%%%%%%%%%%%%%%%%%%%%%%%
\section{Conclusions}
Our main result is that for certain classes of discrete Wilson lines the (K\"ahler) modular symmetry $\Gamma_0(n)$ associated with fixed $\mathbb{Z}_4$- and $\mathbb{Z}_3$-planes is considerably less ($n=16$ and $36$ respectively) than was previously realised. Such planes arise in the $\mathbb{Z}_8$ and $\mathbb{Z}_3 \times \mathbb{Z}_6$ orbifolds respectively. Both orbifolds may be realised by Casimirs acting on $T^4 \oplus T^2$ lattices, with $T^2$ the  $\mathcal{N}=2$ fixed plane. Neither has any fixed $\mathbb{Z}_2$-planes, so the only unfixed moduli are K\"ahler. We also showed that the modular symmetry for $\mathbb{Z}_6$ planes  may be reduced to $\Gamma_0(4)$ for certain classes of discrete Wilson lines.

For the $\mathbb{Z}_8$ case, the allowed modular weights of the Standard Model matter content make
 it  impossible to satisfy the  ``strong'' unification constraint for the supersymmetric Standard Model gauge coupling constants given in eqn (\ref{512}) consistently with the cancellation of modular anomalies. It is, however, possible to satisfy a somewhat weaker constraint that is compatible with the measured values of the gauge coupling strengths $g_a(\mu)$ at low scales $\mu \sim m_Z$. In this case the threshold corrections $\Delta_a$ depend only on the single modulus $T_3$. Using a conjectured form of $\Delta_a$ that possesses the reduced modular symmetry $\Gamma _0(16)$, we have verified that the gap between the observed unification scale $m_X \sim 2 \times 10^{16}$ GeV and the string scale $m_{\rm string} \simeq 3.7 \times 10^{17}$ GeV can now be bridged with a plausible value of $T_3 \sim 1.5$.

For the $\mathbb{Z}_3 \times \mathbb{Z}_6$ case, the situation is more complicated because the orbifold is realised on a $T_1^2 \oplus T_2^2 \oplus T_3^2$ lattice and all three tori are $\mathcal{N}=2$ fixed planes. Thus, all three K\"ahler moduli $T_{1,2,3}$ are unfixed and the threshold corrections depend upon all three; $T^2_1$ and $T^2_2$ are fixed $\mathbb{Z}_6$-planes, while $T^2_3$ is a fixed $\mathbb{Z}_3$-plane. Since there is no reason why any one of the three moduli should dominate, we assume that all three have the same value $T$. In this case, we find that the allowed modular weights {\em are} compatible with the strong unification constraint (for several different values of the allowed anomaly coefficients ${b'_a}^k$). Further, using the conjectured form of the threshold corrections that possesses the full $PSL(2,\mathbb{Z})$ modular symmetry for $T_{1,2}$ and $\Gamma_0(36)$ for $T_3$, we showed that the gap between $m_X$ and $m_{\rm string}$ can be bridged in this case too with  a plausible value of $T \sim 0.9$. Alternatively, if, notwithstanding our prejudice, a highly anisotropic orbifold is realised with $T_3$ the dominant K\"ahler modulus, then again we find that the allowed modular weights are compatible with strong unification also for several different values of the coefficients ${b'_a}^k$. Using the conjectured form of the threshold corrections with modular symmetry $\Gamma_0(9)$ and/or $\Gamma_0(36)$ in this case allows the gap to be bridged with values of $T_3$ in the range $0.94-1.34$.

Our results suggest, but do not prove, that the considerably reduced modular symmetry that arises for large classes of discrete Wilson line backgrounds on the $\mathbb{Z}_8$ and $\mathbb{Z}_3 \times \mathbb{Z}_6$ orbifolds may well allow the observed unification of the gauge coupling strengths to occur  consistently with the  scale naturally arising from the heterotic string and with plausible values of the K\"ahler moduli that determine the overall scale of these  orbifolds. Determining if this suggestion {\em is} true would require the prior calculation of the threshold corrections in the presence of discrete Wilson lines. Since these Wilson lines mix winding numbers with internal lattice momenta, such a calculation is considerably more complicated than when they are absent.

%%%%%%%%%%%%%%%%%%%%%%%%%%%%%%%%%%%%%%%%%%%%%%%%%%%%%%%%%%%%%%%%%%%%%%%%%%%%%%%%%%%%%%%%%%%%%%%%%%%%%%%%%%%%%%%%%%

\end{document}